\documentclass[12pt]{article}
\usepackage{graphicx}

\begin{document}

\title{General formula for finding Mexican hat wavelets by virtue of Dirac's
representation theory and coherent state}
\author{$^{1,2}$Hong-Yi Fan and $^{1}$Hai-Liang Lu \\
$^{1}$Department of Physics, Shanghai Jiao Tong University, \\
Shanghai 200030, China \\
$^{2}$Department of Material Science and Engineering, \\
University of Science and Technology of China,\\
Hefei, Anhui,230026, China} \maketitle
\begin{abstract}
The admissibility condition of a mother wavelet is explored in the context
of quantum optics theory. By virtue of Dirac's representation theory and the
coherent state' property we derive a general formula for finding Mexican hat
wavelets.
\end{abstract}

\section{Introduction}

Wavelet is a `small wave' which is localized in both time and frequency space%
\cite{1,2,3}. It is this unique characteristic that makes wavelets analysis
in some ways superior to Fourier analysis which employs `big waves'
(sinusoid or cosine), e.g., wavelets are particularly useful when processing
data with sharp discontinuities or compressing image data. Mathematically, a
wavelet $\psi $ of the real variable $x$ must satisfy the following
admissibility condition
\begin{equation}
\int_{-\infty }^{\infty }\psi \left( x\right) dx=0,  \label{1}
\end{equation}%
which suggests that $\psi \left( x\right) $ behaves like a wave, and in
contrast to sinusoid, it decreases rapidly to zero as $|x|$ tends to
infinity. The theory of wavelets is concerned with the representation of a
function in terms of a two-parameter family of dilates and translates of a
fixed function, which is usually known as the "mother wavelet". A family of
wavelets $\psi _{\left( \mu ,s\right) }$ ($\mu >0$ is a scaling parameter, $s
$ is a translation parameter, $s\in \mathrm{R)}$ are constructed from the
mother wavelet $\psi $, and the dilated-translated functions are defined as
\begin{equation}
\psi _{\left( \mu ,s\right) }\left( x\right) =\frac{1}{\sqrt{\mu }}\psi
\left( \frac{x-s}{\mu }\right) ,  \label{2}
\end{equation}%
and the wavelet integral transform of a signal function $f\left( x\right)
\in L^{2}\left( \mathrm{R}\right) $ by $\psi _{\left( \mu ,s\right) }$ is
defined by
\begin{equation}
W_{\psi }f\left( \mu ,s\right) =\frac{1}{\sqrt{\mu }}\int_{-\infty }^{\infty
}f\left( x\right) \psi ^{\ast }\left( \frac{x-s}{\mu }\right) dx.  \label{3}
\end{equation}%
The admissibility condition (1) ensures that the inverse transform and
Parseval formula are applicable. When $\psi \left( x\right) $ is an odd
function of $x$, it satisfies (1) obviously. A typical $\psi \left( x\right)
$, which is an even function of $x$, is the Mexican hat wavelet (see fig. 1)
\begin{equation}
\psi _{M}\left( x\right) =\pi ^{-1/4}e^{-x^{2}/2}\left( 1-x^{2}\right) ,
\label{4}
\end{equation}%
satisfying
\begin{equation}
\int_{-\infty }^{\infty }e^{-x^{2}/2}\left( 1-x^{2}\right) dx=0.  \label{5}
\end{equation}%
\begin{figure}[tbp]
\centering \includegraphics[width=2in, height=3in,
angle=-90]{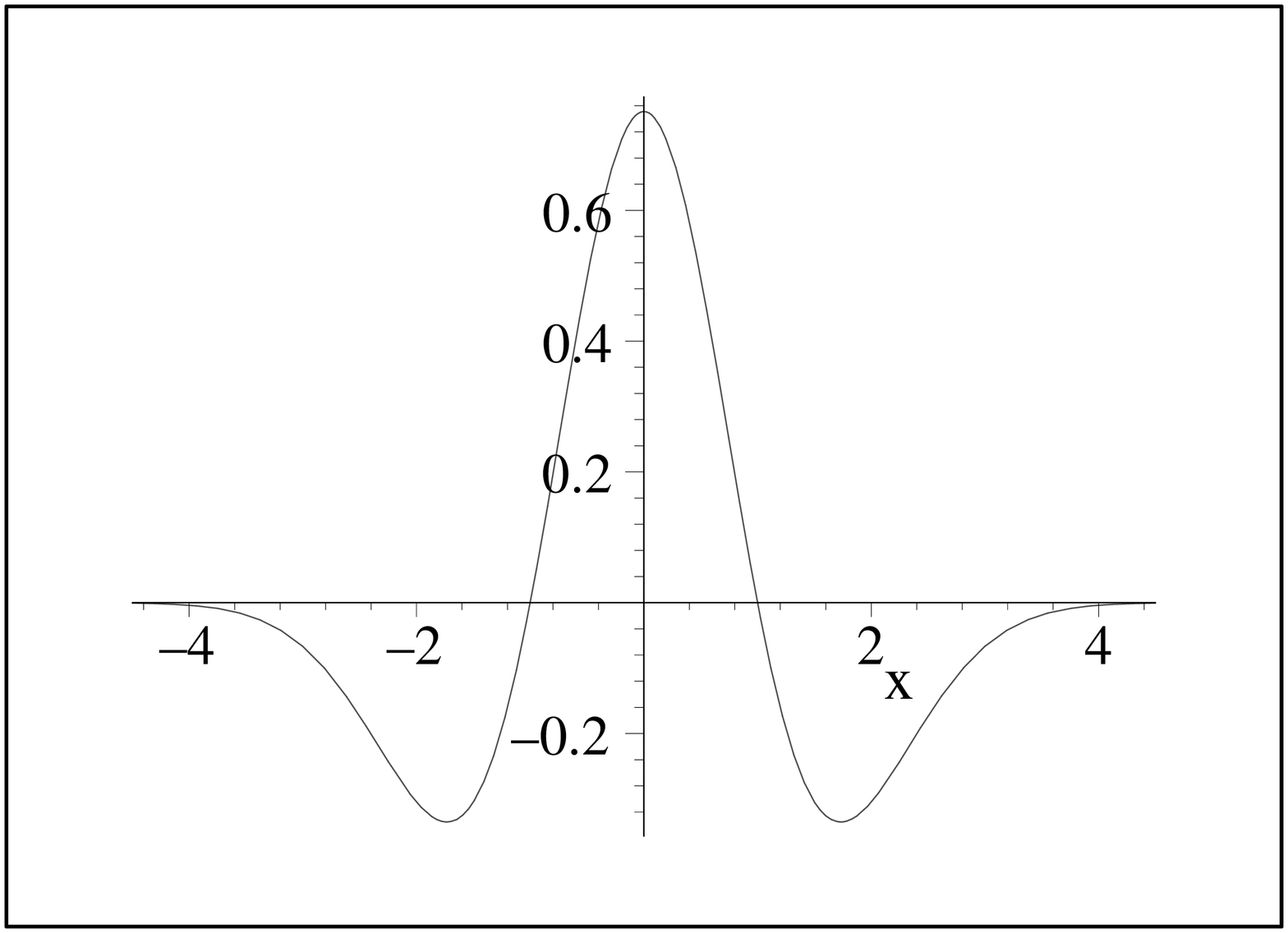}
\caption{{}}
\end{figure}

An important question thus naturally arises: how to find more even functions
which also satisfy (1), i.e. are there a series of even functions which can
be considered as generalized Mexican hat wavelets? To our knowledge, this
question has not been posed and solved in the literature before. In this
letter we shall derive a general formula for finding Mexican hat wavelets by
taking the advantage of Dirac's representation theory and the coherent state
theory in quantum optics, so that more mother wavelets and correspondingly
more wavelet-transformations can be introduced. Dirac's representation
theory is not just a foundation of quantum mechanics theory\cite{4}; it has
its own special features which allow it to be extended to new theoretical
problems. In this Letter we shall show how this theory can help us to
directly derive general formula for finding Mexican hat wavelets.

According to Dirac's representation theory, we can express Eq. (3) as
\begin{equation}
W_{\psi }f\left( \mu ,s\right) =\left\langle \psi \right\vert U\left( \mu
,s\right) \left\vert f\right\rangle ,  \label{6}
\end{equation}%
where $\left\langle \psi \right\vert $ is the state vector corresponding to
the given mother wavelet, $\left\vert f\right\rangle $ is the state to be
transformed, and
\begin{equation}
\frac{1}{\sqrt{\mu }}\int_{-\infty }^{\infty }\left\vert \frac{x-s}{\mu }%
\right\rangle \left\langle x\right\vert dx\equiv U\left( \mu ,s\right)
\label{7}
\end{equation}%
is the squeezing-translating operator\cite{5,6}, $\left\vert x\right\rangle $
is the coordinate eigen-vector of $X$, $X\left\vert x\right\rangle
=x\left\vert x\right\rangle $, which in the Fock space is expressed as
\begin{equation}
\left\vert x\right\rangle =\pi ^{-1/4}\exp \left( -\frac{1}{2}x^{2}+\sqrt{2}%
xa^{\dagger }-\frac{a^{\dagger 2}}{2}\right) |0\rangle ,  \label{8}
\end{equation}%
here $|0\rangle $ is the vacuum state annihilated by the bosonic operator $a$%
, $a|0\rangle =0$, $\left[ a,a^{\dagger }\right] =1$, and $X=\left(
a+a^{\dagger }\right) /\sqrt{2}$. In order to combine wavelet transforms
with transforms of quantum states more tightly and clearly, using the
technique of integration within an ordered product (IWOP)\cite{5} (for a
review see\cite{6,7}) of operators we can directly perform the integral in
(7)%
\begin{eqnarray}
U\left( \mu ,s\right)  &=&\frac{1}{\sqrt{\pi \mu }}\int_{-\infty }^{\infty
}dx:\exp \left[ -\frac{x^{2}}{2}\left( 1+\frac{1}{\mu ^{2}}\right) +\frac{xs%
}{\mu ^{2}}+\sqrt{2}\frac{x-s}{\mu }a^{\dagger }+\sqrt{2}xa-\frac{s^{2}}{%
2\mu ^{2}}-X^{2}\right] :  \nonumber \\
&=&\sqrt{\frac{2\mu }{1+\mu ^{2}}}:\exp \left[ \frac{1}{2\left( 1+\mu
^{2}\right) }\left( \frac{s}{\mu }+\sqrt{2}a^{\dagger }+\sqrt{2}\mu a\right)
^{2}-\sqrt{2}\frac{s}{\mu }a^{\dagger }-\frac{s^{2}}{2\mu ^{2}}-X^{2}\right]
:,  \label{9}
\end{eqnarray}%
where $:$ $:$ denotes normal ordering. Let $\mu =e^{\lambda }$, so sech$%
\lambda =\frac{2\mu }{1+\mu ^{2}}$, tanh$\lambda =\frac{\mu ^{2}-1}{\mu
^{2}+1}$, using the operator identity $e^{ga^{\dagger }a}=:\exp \left[
\left( e^{g}-1\right) a^{\dagger }a\right] :$, Eq. (9) becomes%
\begin{eqnarray}
U\left( \mu ,s\right)  &=&\exp \left[ \frac{-s^{2}}{2\left( 1+\mu
^{2}\right) }-\frac{a^{\dagger 2}}{2}\tanh \lambda -\frac{a^{\dagger }s}{%
\sqrt{2}} sech \lambda \right] \exp \left[ \left( a^{\dagger }a+\frac{1%
}{2}\right) \ln sech \lambda \right]   \nonumber \\
&&\times \exp \left[ \frac{a^{2}}{2}\tanh \lambda
+\frac{sa}{\sqrt{2}} sech\lambda \right] .  \label{10}
\end{eqnarray}%
In particular, when $s=0$, it reduces to the well-known squeezing operator,
\begin{equation}
U\left( \mu ,0\right) =\frac{1}{\sqrt{\mu }}\int_{-\infty }^{\infty
}\left\vert \frac{x}{\mu }\right\rangle \left\langle x\right\vert dx=\exp [%
\frac{\lambda }{2}\left( a^{2}-a^{\dagger 2}\right) .  \label{11}
\end{equation}%
For a review of the squeezed state theory we refer to\cite{8,9,10}.

Now we analyze the condition (1) for mother wavelet in the context of
quantum optics theory. Due to Dirac's representation transformation $\frac{1%
}{\sqrt{2\pi }}\int_{-\infty }^{\infty }\left\vert x\right\rangle
e^{ipx}dx=\left\vert p\right\rangle $, where $\left\vert p\right\rangle $ is
the momentum eigenstate,%
\[
\left\vert p\right\rangle =\pi ^{-1/4}\exp \left( -\frac{p^{2}}{2}+\sqrt{2}%
ipa^{\dagger }+\frac{a^{\dagger 2}}{2}\right) \left\vert 0\right\rangle ,
\]%
we have
\begin{equation}
\frac{1}{\sqrt{2\pi }}\int_{-\infty }^{\infty }\left\vert x\right\rangle
dx=\left\vert p=0\right\rangle ,  \label{12}
\end{equation}%
which can help us to recast the condition (1) into Dirac's ket-bra
formalism,
\begin{equation}
\int_{-\infty }^{\infty }\psi \left( x\right) dx=0\rightarrow \left\langle
p=0\right\vert \left. \psi \right\rangle =0,  \label{13}
\end{equation}%
which indicates that the probability of a measurement of $\left\vert \psi
\right\rangle $ by the projection operator $\left\vert p\right\rangle
\left\langle p\right\vert $ with the value $p=0$ is zero. Now we want to
find such $\left\vert \psi \right\rangle $ that obeys $\left\langle
p=0\right\vert \left. \psi \right\rangle =0$. By considering $a^{\dagger n}/%
\sqrt{n!}\left\vert 0\right\rangle =\left\vert n\right\rangle $ is the
orthogonal basis of Fock representation, without loss of generality, we can
expand $\left\vert \psi \right\rangle $ as
\begin{equation}
\left\vert \psi \right\rangle =G\left( a^{\dagger }\right) \left\vert
0\right\rangle =\sum_{n=0}^{\infty }g_{n}a^{\dagger n}\left\vert
0\right\rangle ,  \label{14}
\end{equation}%
where $g_{n}$ are such chosen as to let $\left\vert \psi \right\rangle $
obey the condition (13). Then using the over-completeness relation of
coherent states (useful representation in quantum optics and can describe
laser\cite{11,12}),%
\begin{equation}
\int \frac{d^{2}z}{\pi }\left\vert z\right\rangle \left\langle z\right\vert
=1,  \label{16}
\end{equation}%
where
\begin{equation}
\left\vert z\right\rangle =\exp \left( za^{\dagger }-z^{\ast }a\right)
\left\vert 0\right\rangle ,  \label{17}
\end{equation}%
$a\left\vert z\right\rangle =z\left\vert z\right\rangle $, and
\begin{equation}
\left\langle p=0\right\vert \left. z\right\rangle =\pi ^{-1/4}\exp \left(
\frac{z^{\ast 2}}{2}\right) ,  \label{18}
\end{equation}%
we have%
\begin{eqnarray}
\left\langle p=0\right\vert \left. \psi \right\rangle &=&\left\langle
p=0\right\vert \int \frac{d^{2}z}{\pi }\left\vert z\right\rangle
\left\langle z\right\vert \sum\limits_{n}g_{n}a^{\dagger n}\left\vert
0\right\rangle  \nonumber \\
&=&\pi ^{-1/4}\sum\limits_{n}g_{n}\int \frac{d^{2}z}{\pi }%
e^{-|z|^{2}}z^{\ast n}\sum\limits_{m}\frac{1}{m!}\left( \frac{z^{2}}{2}%
\right) ^{m}  \nonumber \\
&=&\pi ^{-1/4}\sum\limits_{m}\sum\limits_{n}\frac{1}{m!2^{m}}g_{n}\delta
_{n,2m}n!=\pi ^{-1/4}\sum\limits_{n}\frac{\left( 2n\right) !}{n!2^{n}}%
g_{2n}=0.  \label{19}
\end{eqnarray}%
Then condition (19) provides a general formalism to find the qualified
wavelet. To illustrate the usage of (19), assuming that in (19) $g_{2n}=0$
for $n>3$, so the coefficients of the survived terms should satisfy
\begin{equation}
g_{0}+g_{2}+3g_{4}+15g_{6}=0,  \label{20}
\end{equation}%
and $\left\vert \psi \right\rangle $ becomes
\begin{equation}
\left\vert \psi \right\rangle =\left( g_{0}+g_{2}a^{\dagger
2}+g_{4}a^{\dagger 4}+g_{6}a^{\dagger 6}\right) \left\vert 0\right\rangle .
\label{21}
\end{equation}%
Projecting it onto the coordinate representation, we get the qualified
wavelets%
\begin{eqnarray}
\psi \left( x\right) &=&\left\langle x\right. \left\vert \psi \right\rangle
\nonumber \\
&=&\pi ^{-1/4}e^{-x^{2}/2}\left[ g_{0}+g_{2}\left( 2x^{2}-1\right)
+g_{4}\left( 4x^{4}-12x^{2}+3\right) \right.  \nonumber \\
&&\left. +g_{6}\left( 8x^{6}-60x^{4}+90x^{2}-15\right) \right] ,  \label{22}
\end{eqnarray}%
where we have used
\begin{equation}
\left\langle x\right. \left\vert n\right\rangle =\frac{1}{\sqrt{2^{n}n!\sqrt{%
\pi }}}H_{n}\left( x\right) e^{-x^{2}/2},  \label{23}
\end{equation}%
and the Hermite polynomials' definition%
\begin{equation}
H_{n}\left( x\right) =\left( -1\right) ^{n}e^{x^{2}}\frac{d^{n}}{dx^{n}}%
e^{-x^{2}}.  \label{24}
\end{equation}%
Now we take some examples and depict them in figures.

Case 1: when we take $g_{0}=\frac{1}{2}$, $\ g_{2}=-\frac{1}{2}$ and $%
g_{4}=g_{6}=0$, we immediately obtain the Mexican hat wavelet (see fig. 1)
as in (4). Hence $\frac{1}{2}\left( 1-a^{\dagger 2}\right) \left\vert
0\right\rangle $ is the state vector corresponding to the Mexican hat mother
wavelet.

Case 2: when $g_{0}=-1$, $g_{2}=-2$, $g_{4}=1$ and $g_{6}=0$, we obtain (see
fig. 2)%
\begin{equation}
\psi _{2}\left( x\right) =4\pi ^{-1/4}e^{-x^{2}/2}\left(
x^{4}-4x^{2}+1\right) ,  \label{25}
\end{equation}%
which also satisfies the condition%
\begin{equation}
\int_{-\infty }^{\infty }4\pi ^{-1/4}e^{-x^{2}/2}\left(
x^{4}-4x^{2}+1\right) dx=0.  \label{26}
\end{equation}%
\
\begin{figure}[tbp]
\begin{minipage}[t]{0.5\linewidth}
\centering \includegraphics[width=1.8in, height=2.5in,
angle=-90]{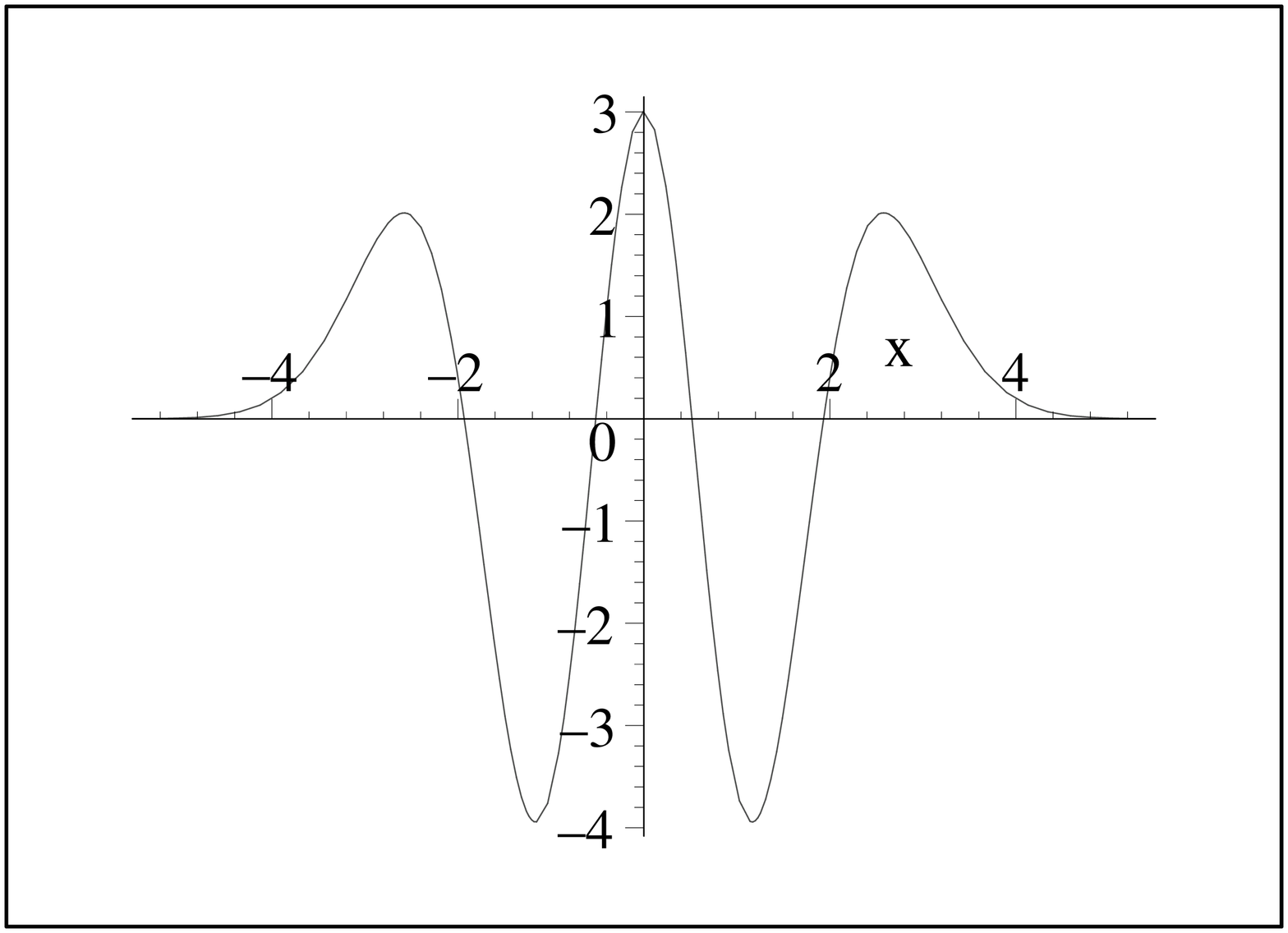} \caption{{}}
\end{minipage}%
\begin{minipage}[t]{0.5\linewidth}
\includegraphics[width=1.8in, height=2.5in,
angle=-90]{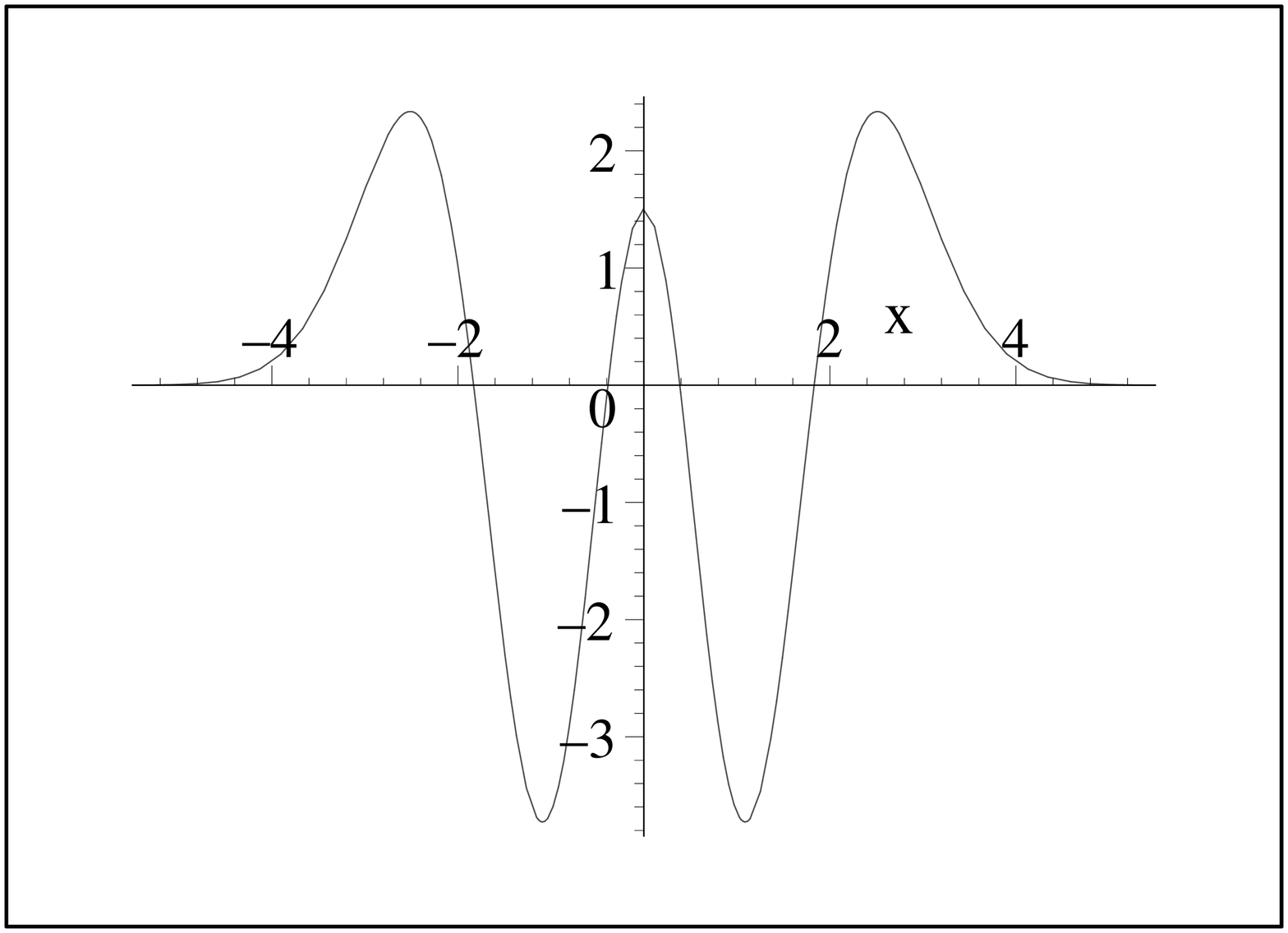} \caption{{}}
\end{minipage}
\end{figure}
Note that when $g_{0}=-2$, $g_{2}=-1$, $g_{4}=1$ and $g_{6}=0$, we obtain a
slightly different wavelet (see fig. 3). Therefore, as long as the
parameters conforms to condition (19), we can adjust their values to control
the shape of the wavelet.

Case 3: when $g_{0}=1$, $g_{2}=2$, $g_{4}=4$ and $g_{6}=-1$, we get (see
fig. 4)%
\begin{equation}
\psi _{3}\left( x\right) =\pi ^{-1/4}e^{-x^{2}/2}\left(
-8x^{6}+76x^{4}-134x^{2}+26\right) ,  \label{27}
\end{equation}%
and
\begin{equation}
\int_{-\infty }^{\infty }\pi ^{-1/4}e^{-x^{2}/2}\left(
-8x^{6}+76x^{4}-134x^{2}+26\right) dx=0.  \label{28}
\end{equation}%
\begin{figure}[tbp]
\centering \includegraphics[width=1.8in, height=2.5in,
angle=-90]{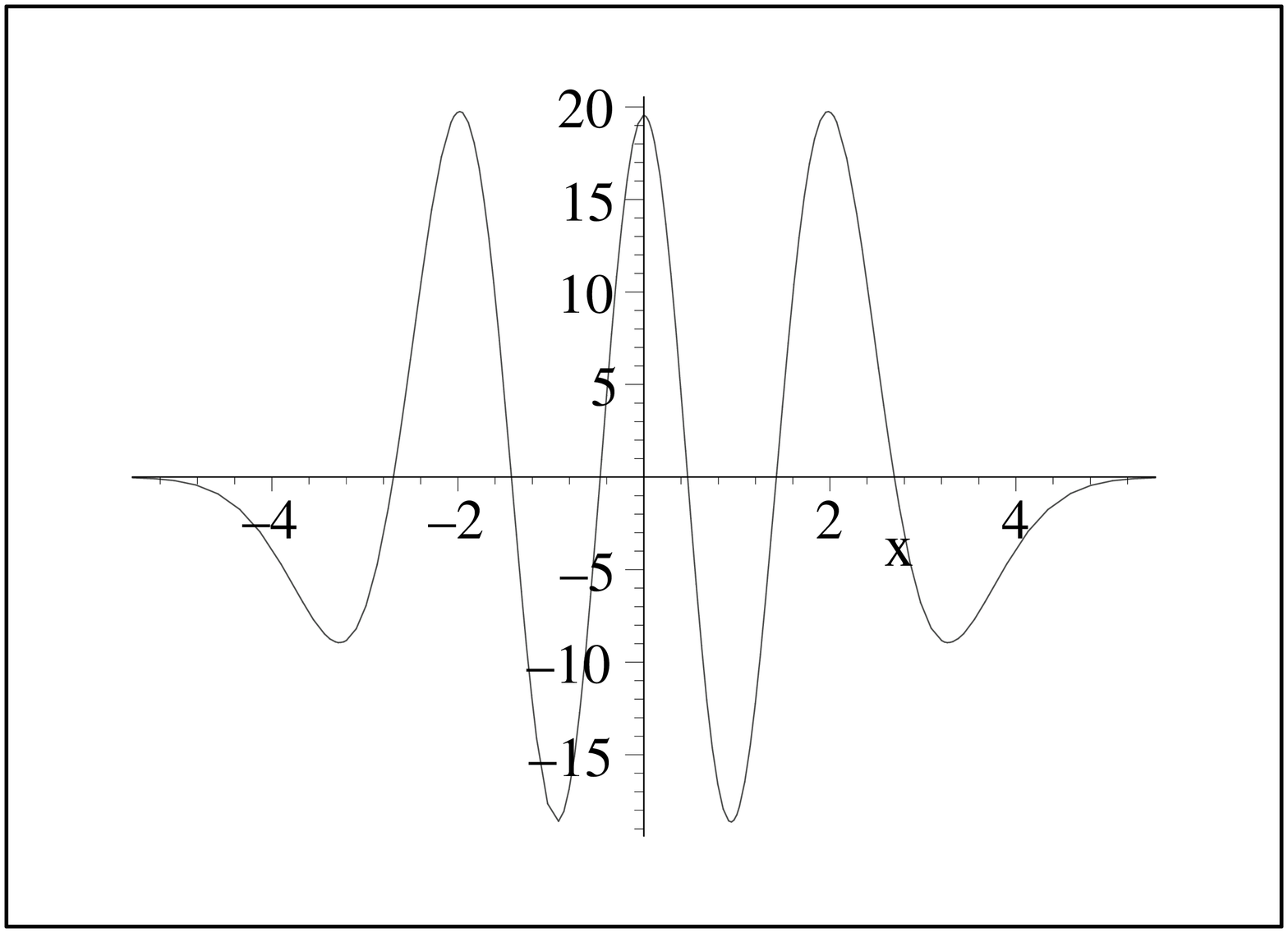}
\caption{{}}
\end{figure}
From these figures it is observed that the number of the crossing points of
the curve at the $x$-axis is equal to the highest power of the wavelet
function.

In summary, by converting wavelets and its admissibility condition into the
framework of Dirac's ket-bra formalism and using the coherent state's
well-behaved properties we have derived the general formula for composing
Mexican hat wavelets, based on which more qualified wavelets can be found
and more wavelet transformations can be defined. This work again shows the
powerfulness of Dirac's representation theory.

\end{document}